\documentclass[         
aps,                    
prl,                    
showpacs,               
nofootinbib,            
twocolumn,             %
showkeys,               %
preprintnumbers,        %
amsmath,               %
amssymb,               %
floatfix]               
{revtex4}               
\usepackage{graphicx,longtable}
\usepackage{color}

\begin{document}

\title{A dynamical collective calculation of supernova neutrino signals}

\author{J\'er\^ome Gava, James Kneller, Cristina Volpe}
\email{gava@ipno.in2p3.fr,kneller@ipno.in2p3.fr,volpe@ipno.in2p3.fr}
\affiliation{Institut de Physique Nucl\'eaire, F-91406 Orsay cedex, CNRS/IN2P3 and University of Paris-XI, France}

\author{G.C. McLaughlin}
\email{gcmclaug@ncsu.edu}
\affiliation{Department of Physics, North Carolina State University,
Raleigh, North Carolina 27695-8202}

\begin{abstract}
We present the first calculations with three flavors of collective and 
shock wave effects for neutrino propagation in core-collapse supernovae using 
hydroynamical density profiles and the S matrix formalism. 
We explore the interplay between the neutrino-neutrino interaction and
the effects of multiple resonances upon the time signal of
positrons in supernova observatories. A specific signature is found for the inverted hierarchy and a large third 
neutrino mixing angle and we predict, in this case, a dearth of lower energy positrons in Cherenkov detectors midway through the 
neutrino signal and the simultaneous revelation of valuable information about the original fluxes. 
We show that this feature is also observable with current generation neutrino detectors at the level of
several sigmas.
\end{abstract}


\date{\today}

 \maketitle


\noindent
{\it Introduction}
Core-collapse supernovae are rare but represent impressive
violent phenomena in the Universe capable of emitting 10$^{53}$ ergs as
neutrinos of all flavors
in a short burst of about 10 seconds. While such neutrinos were first
observed in 1987, current observatories or 
next generation detectors, could provide us with the
luminosity curve from a future (extra)galactic explosion and/or the exciting
observation of relic neutrinos from past supernovae.
Untangling the information from such neutrino signals will represent
a challenging task, since
the information from the explosion phenomenon and the neutrino
properties -- e.g. the neutrino hierarchy and the third neutrino mixing
angle -- is intertwined. 

Impressive progress has been achieved in the last few years in our
understanding of how neutrinos propagate in supernovae, fundamentally modifying the
standard MSW effect paradigm. This evolution is due 
to the substantial progress made in the calculations which now  either include
neutrino-neutrino interactions or use profiles which contain multiple resonances.
Pantaleone first pointed out that the neutrino-neutrino interactions lead to an off-diagonal
coupling  \cite{Pantaleone:1992eq}, and subsequent
numerical calculations showed that new phenomena arose when such terms
are included in the neutrino evolution equations \cite{Samuel:1993uw}. 
It is now well
established that neutrinos evolve collectively through the synchronized and bipolar regimes eventually exhibiting a spectral 
split (see e.g. 
\cite{Duan:2005cp,Pastor:2001iu,Hannestad:2006nj,Raffelt:2007xt,Dasgupta:2007ws}).
The phenomenological implications are important,
especially for the inverted hierarchy case, because electron anti-neutrinos get a ``hot'' 
spectrum, independently of the
smallness of $\theta_{13}$. A second swapping of the anti-neutrino fluxes
may also occur when they reach the MSW resonance region, depending on the
hierarchy and of the size of $\theta_{13}$. Recent works have
pointed out that the dynamics of the explosion can alter the high density resonance \cite{Schirato:2002tg} and have found  that the
neutrinos might encounter the resonance density more than once
(multiple resonances) \cite{Tomas:2004gr,Fogli:2004ff,Kneller:2005hf}. 
Rapid oscillations in the (anti)neutrino
propabilities might appear because of the phase differences between two or more resonances \cite{Dasgupta:2005wn,Kneller:2007kg}. 

This paper presents the first numerical calculation with three flavors to include both the
neutrino-neutrino interaction and dynamic MSW effects using matched density profiles and 
correctly putting them together using S matrices rather than probabilities. 
We focus upon the
results of the anti-neutrino time signal. Signatures 
pinpointing to the hierarchy and possible $\theta_{13}$ values are shown, even for values beyond 
the proposed reach of future experiments on Earth. For large $\theta_{13}$ and inverted hierarchy
we predict that the positron event rate possesses a characteristic feature due to the passage of the shock. Such a feature is observable in current
and future large-size neutrino observatories.

\noindent
{\it Theoretical framework}
We calculate the three flavor neutrino evolution through matter in two steps.
First, we evolve 
the neutrino wavefunctions up to some radius using supernova density profiles at different times during the supernova explosion, as done in \cite{Gava:2008rp} for one static density profile. This calculation includes the neutrino coupling both to
matter with loop corrections (i.e. $V_{\mu\tau}$) and to neutrino themselves. For the latter we use the single-angle
approximation, i.e. we assume that neutrinos are essentially emitted with one
angle. Such an assumption accounts rather well both qualitatively and
quantitatively for the neutrino collective effects \cite{Duan:2006jv,Fogli:2007bk}, 
even though in some cases decoherence in a full multi-angle description might appear (see e.g. \cite{EstebanPretel:2007ec}). 
The density profile used is a dynamic inverse power-law. 
The second step is to determine the exact neutrino evolution through the rest of the supernova mantle by solving the evolution operator
equations as described in \cite{Kneller} which is a 3 flavor generalization of \cite{Kneller:2005hf}. The 1D density profiles used 
are taken from the hydrodynamical calculations described in 
\cite{Kneller:2007kg} and include both a front and a reverse shock. 
These profiles are matched to the profiles used in the first step. The full results are then sewn together using the S matrices rather 
than probabilities c.f. \cite{Kneller:2007kg,Lunardini:2007vn}.
Finally, the flux on Earth is calculated taking into account
decoherence \cite{Dighe:1999bi} but not Earth matter
\cite{dasgupta-2008-101} which would be relevant if the supernova were shadowed. 

\noindent
{\it Numerical results}
Our main goal is to explore the neutrino time signal in an observatory, depending on the yet unknown neutrino parameters, and see if we can exploit a combination of the neutrino-neutrino interaction and shock wave effects to get clues on important open issues. 
We take as an example the electron anti-neutrino scattering on protons which is the dominant channel in Cherenkov and scintillator detectors. The results we present are obtained with the best fit oscillation parameters, i.e. $\Delta m^2_{12}= 8 \times 10^{-5}$eV$^2$, sin$^2 2\theta_{12}=0.83$ and
$|\Delta m^2_{23}|= 3 \times 10^{-3}$eV$^2$, sin$^2 2\theta_{23}=1$ for
the solar and atmospheric differences of the mass squares and
mixings, respectively \cite{Amsler:2008zz}. The Dirac phase is taken to be zero: no effects show up when the muon and tau luminosities are taken equal \cite{Balantekin:2007es}; while a few percent modification can appear due to $V_{\mu \tau}$ and of non-linear effects \cite{Gava:2008rp}.
\begin{figure}[t]
\includegraphics[scale=0.3,angle=-90]{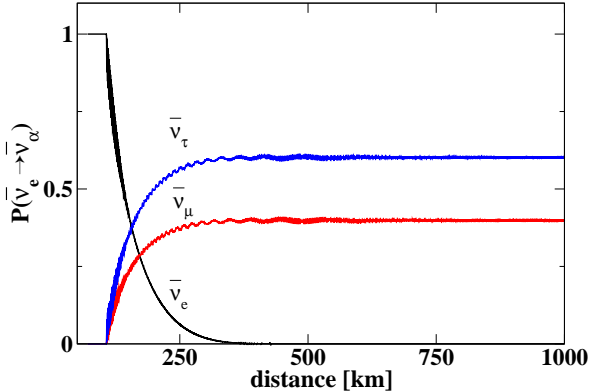}
\caption{Antineutrino oscillation probabilities, as a function of the distance from the neutron-star surface, including the $\nu\nu$ interaction and the $V_{\mu \tau}$. The 
curves correspond to anti-neutrinos 
having 15 MeV energy, the hierarchy is inverted and $\theta_{13}$ is large. }
\label{fig:prob}
\end{figure}

\begin{figure}[t]
\includegraphics[scale=0.3,angle=0.]{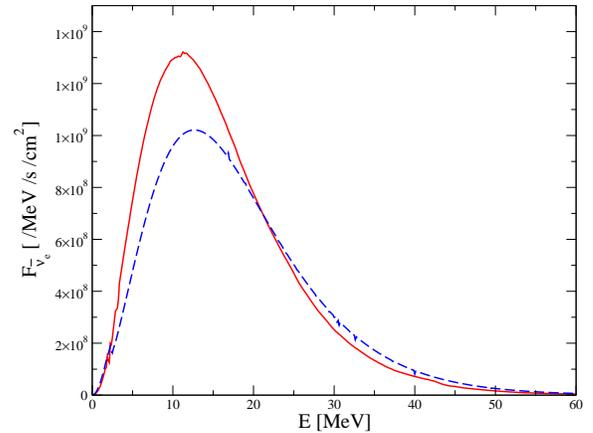}
\caption{Electron anti-neutrino fluxes on Earth, in the case of
adiabatic (``cold'' spectrum, solid) and non-adiabatic (``hot''
spectrum, dashed) conversions in the star. The time is 1 s.}
\label{fig:earth}
\end{figure}

\begin{figure}[t]
\hspace{.1cm}
\includegraphics[scale=0.3,angle=-90.]{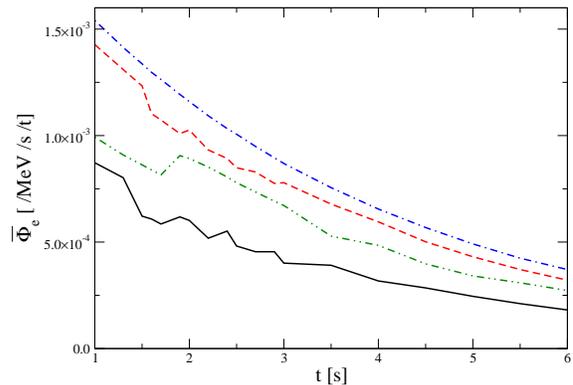}
\caption{Positron time signal per unit tonne in a detector, 
for a galactic explosion at 10 kpc. 
The case of inverted hierarchy and large $\theta_{13}$
 is considered. 
The results are obtained for 
10 (solid), 15 (dashed), 19 (dash-dotted) and 29 
(dot-dot-dashed line) MeV positron energies.}
\label{fig:timesignal}
\end{figure}
One of the important open questions is the hierarchy, since the sign of $\Delta m^2_{23}$ can be positive (normal) or negative (inverted hierarchy).
The value of the third neutrino mixing angle is another issue, particularly crucial for the search of CP violation in the lepton sector. We take here two possible values for $\theta_{13}$, a large (sin$^2 \theta_{13}= 10^{-4}$)  and a small (sin$^2 \theta_{13}= 10^{-8}$). Note that the results corresponding to the large value are emblematic of the whole range sin$^2 \theta_{13}> 10^{-4}$ up to the present experimental Chooz limit. They correspond to the case of the adiabatic conversion at the high density resonance \cite{Lunardini:2003eh}. The other value is chosen as an example of the non-adiabatic regime. Note that sin$^2 2\theta_{13}=10^{-4}-10^{-5}$ are the smallest values that can be reached in accelerator experiments even with very long-term projects  \cite{Volpe:2006in}. 

For the neutrino luminosities at the neutrino-sphere we make the usual
assumption of equipartition of energies among all neutrino flavors and
that they decrease as $L_{\nu}=L_{\nu_0} \times
\exp(-t/\tau)$ with $L_{\nu_0}=10^{52}$ erg $\cdot$ s$^{-1}$ 
and $\tau=3.5$ s. Equal luminosities 
are appropriate for the cooling phase of the neutrino signal upon which we are focusing; during the accretion phase 
the $\nu_e$ and $\bar{\nu}_e$ luminosities are substantially brighter than the $\nu_x$ \cite{Liebendoerfer:2003es}. We consider the hierarchy i.e. $ \langle E_{\nu_e} \rangle < \langle E_{\bar{\nu}_e} \rangle < \langle E_{\nu_{x}} \rangle$ with typical values of 12, 15 and 18 MeV respectively. 

Figure \ref{fig:prob} shows the evolution of probabilities for an anti-neutrino that 
is produced at the neutrino-sphere as an electron-type. One recognizes 
the synchronization region and the rapid bi-polar oscillations  near
the neutrino-sphere. 
The electron anti-neutrino probability becomes very small 
implying that due to the $\nu\nu$ interaction, the corresponding 
fluxes have swapped with the $\nu_{\mu}, \nu_{\tau}$ neutrino fluxes and has turned ``hot'' at 
this point in their propagation. 

To help in understanding the results, let us now consider a supernova explosion located at 10 kpc from Earth.  At the early times (t $\lesssim$ 1 s), for the inverted hierarchy and large $\theta_{13}$, anti-neutrinos undergo an adiabatic MSW resonance and have a ``cold'' spectrum on Earth\footnote{Note that the spectra mix slightly due to the $\theta_{12}$ rotation at the L resonance.} (Figure \ref{fig:earth}).  When the shock wave passes through the MSW high resonance region, 
important modifications of the neutrino fluxes occur. The presence of the shock wave renders the neutrino flavor conversion 
non-adiabatic. Therefore the neutrino spectra on Earth remains ``hot'' (Figure \ref{fig:earth}). Note that the adiabatic and non-adiabatic spectra will cross at some energy, which in this case, is $E_{\nu}=20$~MeV.

Our predictions for the positron time signal associated to inverse
beta-decay in a detector are shown in Figure \ref{fig:timesignal}. Let
us consider the case of inverted hierarchy and large third neutrino mixing angle.  At early times, the conversion is adiabatic and neutrinos with less than 20 MeV will produce a number of positrons determined by the ``cold'' spectrum. This number will decrease when the shock wave renders the resonance conversion non-adiabatic and the neutrinos reach the Earth with a ``hot'' spectrum. This will show up in the positron time signal as a dip. Its depth is energy dependent. (Note that if the $\nu\nu$ interaction is absent, the time signal would show a bump instead, as discussed in \cite{Kneller:2007kg}.) 
For energies larger than 20 MeV, since the relative number flux  of ``hot'' and ``cold'' spectra interchange, the dip turns into a bump. 

\begin{figure}
\includegraphics[scale=0.3,angle=0]{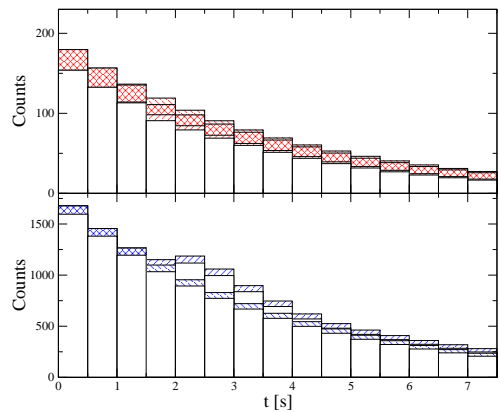}\hspace{.2cm}
\caption{Positron time signal in Super-Kamiokande, taking  0.5 s time bins, and two energy bins : 10 to 19 MeV (upper figure) and above 25 MeV (lower figure). An exponential decay calibrated on the first two time bins is shown for comparison.}
\label{fig:superK}
\end{figure}

Figure \ref{fig:superK} presents our prediction for Super-Kamiokande (22.5 ktons) for which about ten thousand events are expected. If one takes 0.5 s time bins, and a lower energy bin between 10 and 19 MeV and a higher one, above 25 MeV,  one can see that  a bump (dip) can be distinguished by the pure exponential behaviour at 3.5 (1) sigma for the upper (lower) energy range. Obviously such a signature could be observed with exquisit precision if a next-generation large-size detector such as MEMPHYS (UNO, Hyper-K) is considered \cite{Autiero:2007zj}. The large number of events would allow to use small energy bins and follow smoothly the transition from bump to dip. 

\begin{figure}[t]
\includegraphics[scale=0.3,angle=0.]{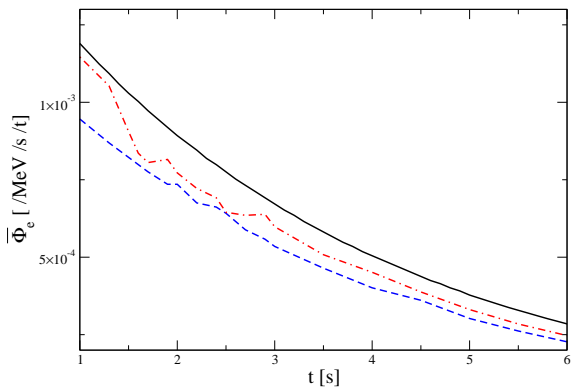}
\caption{Positron time signal associated to inverse beta-decay in a detector,  for a 12 MeV positron. The results correspond to a normal hierarchy (solid) or an inverted hierarchy with large (dot-dashed) or small (dashed) third neutrino mixing angle. The presence of a dip is typical of the whole range $\sin^2 \theta_{13}=0.1-10^{-4}$.}
\label{fig:timesignal2}
\end{figure}

Finally, let us discuss the sensitivity to the hierarchy and upon the third neutrino mixing angle (Figure \ref{fig:timesignal2}). 
If the hierarchy is inverted and $\theta_{13}$ is large the positron
signal as a function of time presents a dip as discussed above. If
$\theta_{13}$ is small, no flavor conversion occurs in the high
density resonance region and the electron anti-neutrino fluxes remain ``hot''. 
This leads to simple exponential decay, in the positron time signal.
On the other hand if the hierarchy is normal the anti-neutrino fluxes
do not change traversing the whole star  and remain ``cold'' during the explosion. 
In this case the positron time signal again is an exponential decay but more positrons 
are produced compared to the inverted hierarchy case.

\noindent
{\it Conclusions}
We have performed the first calculation to include 
the most recent developments of neutrino propagation in dense media such as core-collapse supernovae. Our numerical results include the neutrino-neutrino interaction on one hand and use evolving density profiles from realistic simulations which include multiple resonances on the other. 
Our results show that the interplay between the neutrino-neutrino
interaction and the shock wave can, indeed, be understood in terms of
spectral swapping.
We have explored the positron time signal related to electron anti-neutrino on proton scattering, the dominant detection channel in Cherenkov and scintillator detectors. We have shown that the positron event rate
possesses a characteristic time signal that depends upon the neutrino hierarchy and third neutrino mixing angle. For an inverted hierarchy and a large third neutrino mixing angle the event rate is found to decrease (increase) midway through the supernova neutrino signal for low (high) neutrino energies. 
Such a signature is observable in current and future large-size neutrino detectors at present under study.
Note that this general prediction agrees qualitatively with the observed gap of low energy events in the supernova 1987A data \cite{Hirata:1987hu,Bionta:1987qt,Alekseev:1987JETPL..45..589A} although, with so few observations, an emission model with no shock effects is marginally compatible \cite{Lattimer:1989zz}.
The signature relies upon: a) that the proto-neutron star is 
brighter in $\bar{\nu}_e$ and $\bar{\nu}_x$ at low energies, b) that the hierarchy is inverted so that collective effects swap the 
$\bar{\nu}_e$ and $\bar{\nu}_x$ spectra prior to the H resonance, c)
that neutrino propagation through the progenitor profile is adiabatic, d) the shock reaches the H resonances while the supernova is still luminous. If, for example, the luminosities 
of $\bar{\nu}_e$ than $\bar{\nu}_x$ at low energy were comparable then
there would be little decrease in the positron event rate midway through 
the signal but, on the flip side, a greater difference in luminosity exagerates the event rate decrease. 
This signature might be robust in the presence of
turbulence, although further investigation is required. 
The reasoning is that the density profiles used 
here can be thought of as equivalent 
to the `average' profiles of Fogli {\it et al.} \cite{Fogli:2006JCAP...06..012F}. 
When these authors added turbulence 
to the post shock region of their profiles they found that the size of the shock 
effects were muted but, more importantly, they were not removed. Finally the 
observation of the cross-over energy provides valuable information above the original neutrino fluxes.  By measuring different positron energies in a possible future galactic core-collapse supernova explosion one 
might learn if the third neutrino mixing angle is within (smaller) the window of achievability of present  (future) terrestrial experiments. 

J.G., J.K. and C.V.  acknowledge support from Project No. ANR-05-JCJC-0023, G.C.M. from the Department of Energy under contract DE-FG02-02ER41216.

\textit{}

\end{document}